# The sphaleron energy for SU(2)-Higgs from cooling

Margarita García Pérez[a] and Pierre van Baal [a] [*]

[a]Instituut-Lorentz for Theoretical Physics,
University of Leiden, PO Box 9506,
NL-2300 RA Leiden, The Netherlands.

The cooling algorithm for saddle points presented in ref. [1] is generalized to obtain static classical solutions of the SU(2)-Higgs field theory in the limit of infinite Higgs self-coupling. The sphaleron energy obtained via this algorithm is $\mathcal{E}_{\rm sph} = 5.08(7)\, M_W/\alpha_W$ to be compared with 5.0707 obtained in the variational approach [2].

The spherically symmetric sphaleron solutions for the SU(2)-Higgs field theory, obtained by a variational analysis, have been known for quite some time [3,4]. Above $M_H = 12 M_W$ the sphaleron undergoes a series of bifurcations [2], acquiring at each bifurcation an additional negative mode, while a new solution, a so-called deformed sphaleron splits off. For infinite $\lambda$, where the model is identical to the gauged non-linear sigma model, there is an infinite number of solutions ranging in energy from $5.41 M_W/\alpha_W$ to the energy of the lowest deformed sphaleron $5.07 M_W/\alpha_W$, which has only one negative mode with $\omega^2 = -4.714 M_W^2$ (the number of unstable modes increases with increasing energy). At infinite Higgs self-coupling these solutions are related to the so-called electro-weak skyrmions [5]. Generalizing the pure gauge cooling algorithm for saddle points, we present the lattice results for the energy of this lowest deformed sphaleron, henceforth called the sphaleron (using the more restrictive definition that requires the existence of precisely one unstable mode). We view it as a useful check for the efficiency of the algorithm, but one might envisage useful applications for the computation of sphaleron transition rates [6].

## 1. The algorithm

The dynamical variables for the SU(2)-Higgs model on the lattice are the gauge group variables $V_\mu(x)$, defined on the link that runs fron $x$ to $x+\hat{\mu}$, and the Higgs field in the fundamental representation of SU(2). For infinite self-coupling in the Higgs sector, the length of the Higgs doublet is frozen and can be chosen equal to unity. The Higgs field can in this case be represented by a SU(2) matrix $\sigma(x)$, which is associated to the gauge degree of freedom and can be reabsorbed into the links via the change of variables [7]

$$U_\mu(x) = \sigma(x) V_\mu(x) \sigma(x+\mu) \quad . \tag{1}$$

The lattice action is ($U_\mu(x) = \underset{x\quad\mu}{\bullet\!\!\longrightarrow}$)

$$S = \frac{a^n}{g^2 a^4}\left\{\sum_{x,\mu,\nu}\mathrm{Tr}\left(1 - {}^{\nu}\!\!\underset{x\quad\mu}{\square}\right) + \kappa\sum_{x,\mu}\mathrm{Tr}\left(1 - {}^{\mu}\!\!\underset{x}{\updownarrow}\right)\right\} . \tag{2}$$

For $n=4$ ($n=3$) the continuum action (energy) functional is recovered by setting $\kappa = 2 M_W^2 a^2$, with $a$ the lattice spacing. In what follows we will restrict to the $n=3$ case, but the method is applicable in any number of dimensions.

Designing a cooling algorithm for this model goes exactly as for the pure gauge theory. We construct the functional $\hat{S}$ by squaring the equations of motion for $S$

$$\hat{S} = \frac{a^n}{g^2 a^6}\sum_{x,\mu}\mathrm{Tr}\left(\tilde{U}_\mu(x)\tilde{U}_\mu^\dagger(x) - (U_\mu(x)\tilde{U}_\mu^\dagger(x))^2\right) , \tag{3}$$

where

$$\tilde{U}_\mu(x) = \frac{\kappa}{2} + \sum_{\nu\neq\mu}\left( {}^{\nu}\!\!\underset{x}{\sqcap}^{\mu} + {}^{x}_{\nu}\!\!\underset{\mu}{\sqcup} \right) \quad , \tag{4}$$

The equations of motion for $S$ are solved by

$$U_\mu(x) = \pm\tilde{U}_\mu(x)/\|\tilde{U}_\mu(x)\| \quad . \tag{5}$$

[*]Based on the talk presented by the first author at Lat'94 (Bielefeld, Sept. 1994)

The $+$ sign is to be taken in order for the solution to have a smooth continuum limit. Minimizing $S$ amounts to iteratively replacing $U_\mu(x)$ by eq. (5).

To minimize $\hat{S}$ with respect to a single link variable would require us to solve an eighth order polynomial in $\|U_\mu(x)\|$ at each iteration. An alternative algorithm [1] can be designed by noting that $\hat{S}$ will always be lowered under the update

$$U'_\mu(x) = \frac{M(U_\mu(x)) - W_\mu(x)}{\|M(U_\mu(x)) - W_\mu(x)\|} \quad , \qquad (6)$$

with ($V^\alpha_\mu(x)$ are the $2(n-1)$ staples in eq. (4))

$$M(U_\mu(x)) \equiv 2\ \mathrm{Tr}\left(U_\mu(x)\tilde{U}^\dagger_\mu(x)\right)\tilde{U}_\mu(x) + 6\sum_\alpha \mathrm{Tr}\left(U_\mu(x)V^\alpha_\mu(x)^\dagger\right)V^\alpha_\mu(x) \quad , \qquad (7)$$

and

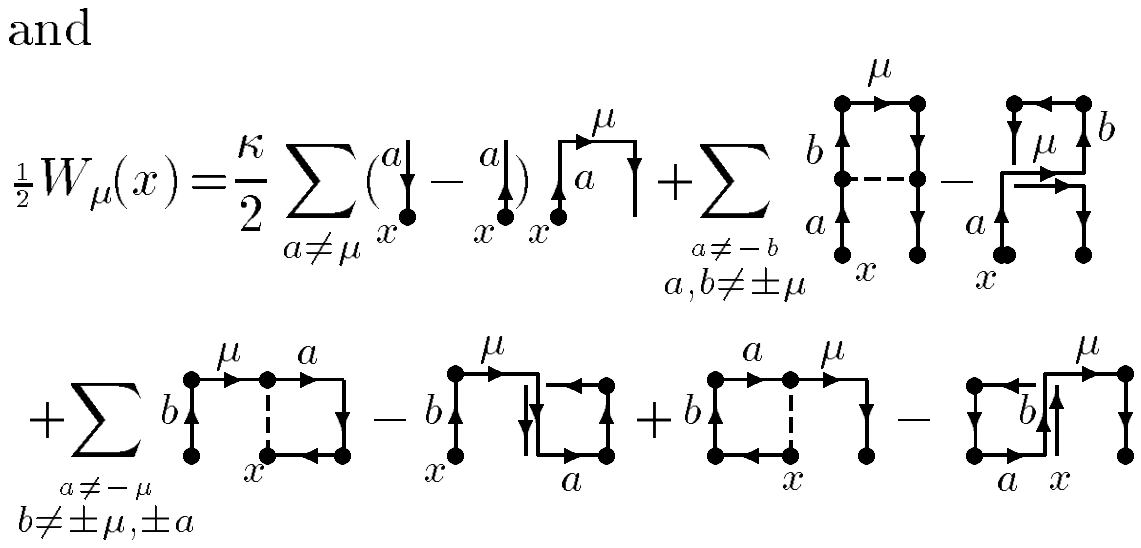

with the unit vector $\hat{a} \in \{\pm\hat{1}, \cdots, \pm\hat{n}\}$. We give the explicit form for the term proportional to $\kappa$, refering for the other terms to ref. [1],

$$\kappa\sum_{a\neq\mu}(I - U_a^2(x))\,U_\mu(x+\hat{a})\,U_a^\dagger(x+\hat{\mu}) \quad , \qquad (8)$$

with the convention $U_{-a}(x) \equiv U_a^\dagger(x-\hat{a})$. This algorithm will have any extremum of the energy functional, even lattice dislocations, as a minimum. To avoid being trapped in one of these dislocations we first bring $S$ down with eq. (5) to a smooth configuration. We then use eq. (6). When this does not lower $\hat{S}$ further some updates with eq. (5) often help (see ref. [1] for details).

## 2. Results

Before presenting the results it is useful to clarify a few points. We are looking for infinite volume solutions. To minimize boundary effects due to the finite lattice size, the typical correlation length of the system has to be much smaller than the size of the box $(aM_W)^{-1} = \sqrt{2/\kappa} \ll N$, on a lattice of size $N^3$. This gives, for fixed $N$, a lower bound on the values of $\kappa$ to be considered. The simulations were performed for $N = 8, 12$ and 16, imposing $N\sqrt{\kappa/2} \geq 2.5$. For smaller $\kappa$ values it turns out that the electro-weak sphaleron developes additional unstable modes. This is not due to a bifurcation, but due to the crossing in energy of two widely separated critical points. The other critical point responsible for this is the finite volume sphaleron, constructed for $\kappa = 0$ in ref. [1]. (It is interesting to note that this is not forbidden by Morse theory, which measures the global topology. Only if the Morse functional - i.e. the energy functional - has no degenerate critical points, bifurcation is implied when additional unstable modes appear).

On the other hand, for large $\kappa$ at fixed $N$, the energy density becomes highly peaked over a few lattice spacings, giving large lattice artefacts. One would like the correlation length to be much bigger than the lattice spacing, $\sqrt{2/\kappa} \gg 1$. In the continuum there is an infinite set of sphaleron-like solutions in a very small range of energies, with no bound on the number of unstable modes. On the lattice there can, however, not be more unstable modes than the finite number of degrees of freedom. The way the solutions bifurcate now also depends on the lattice spacing. The largest lattice spacing we can allow is the one where our sphaleron acquires more than one unstable mode. This can be monitored by computing the Hessian, which is only practical for $N = 8$. From this we determined the bound that $N\sqrt{\kappa/2} \leq 3.2$ (i.e. the correlation length has to be slightly bigger than two lattice spacings).

The way we obtained the required configurations was by starting at $N = 8$ with the links at the boundary frozen to unity. This lifts the energy of the finite volume sphaleron considerably, such that the value of $\kappa$ at this stage may even be below the value we quoted above. Also, lattice artefacts cause the breakdown of translational invariance on a periodic lattice, thereby generating spurious saddle points with up to three extra unstable modes. It has the additional advantage of centering the energy profile in a maximally symmetric way, whereby one minimizes the lattice artefacts. We then release the frozen boundary condition and compute after cooling the Hessian to verify that we have one unstable mode only (at this stage $\kappa$ has to be brought to the window of values discussed above). For $N = 12$ and 16 the initial configurations were generated from the one at $N = 8$, by embedding it in the large lattice (links parallel to the boundary remain constant and those perpendicular to the boundary are put to unity) and adjusting $\kappa$ such that the physical volume is at least as large as for $N = 8$. The results for the sphaleron energies, as well as the negative eigenvalue for the $N = 8$ Hessian are presented in the table below. The values of $\hat{S}$ are not larger than $10^{-4}M_W^3/\alpha_W$ and even considerably smaller for the smaller lattices.

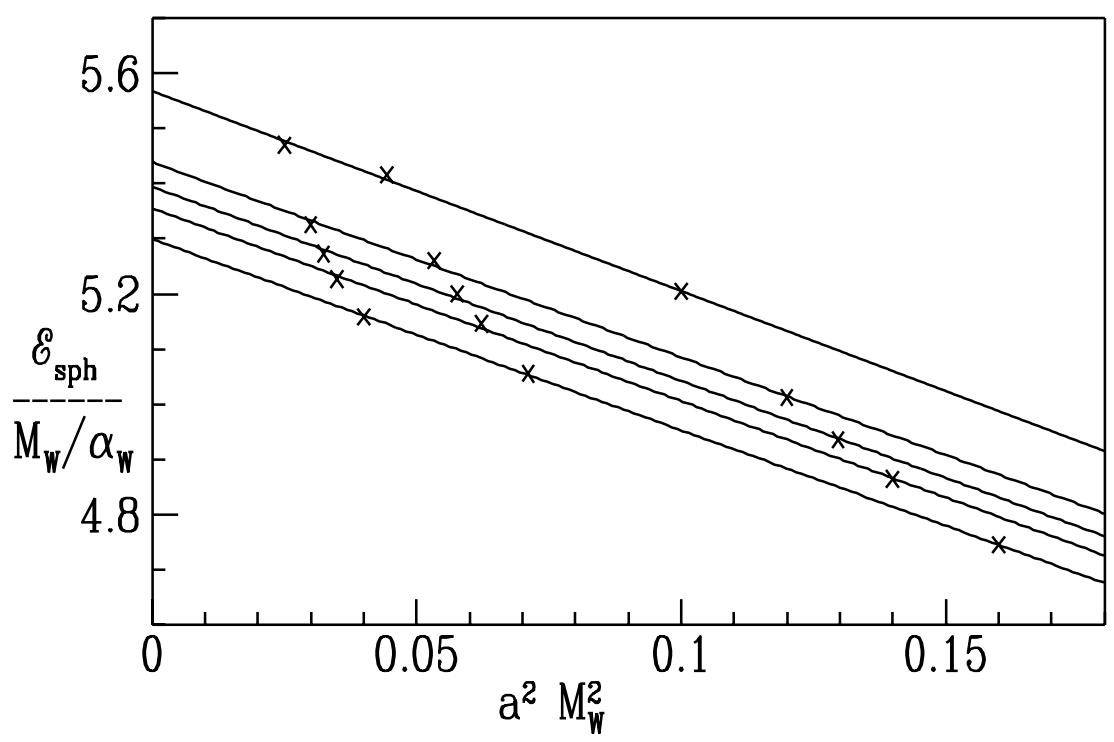

Figure 1. $\mathcal{E}_{\rm sph}$ as a function of $a^2M_W^2$ fitted for fixed physical volume to a linear function of $a^2$.

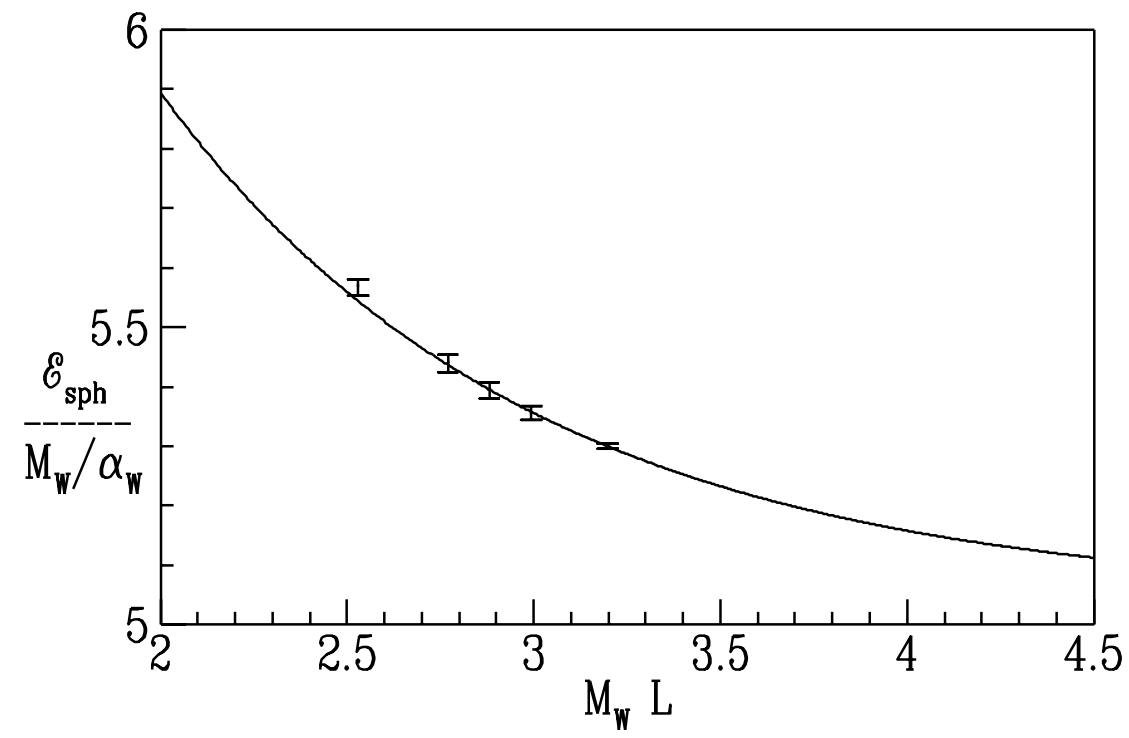

Figure 2. Continuum extrapolated values for $\mathcal{E}_{\rm sph}$ as function of the physical volume $M_WL$.

| $N\sqrt{\frac{\kappa}{2}}$ | $\frac{\mathcal{E}_{\rm sph}}{M_W/\alpha_W}$ | | | $-\frac{\omega^2}{M_W^2}$ |
|---|---|---|---|---|
| | $N=8$ | $N=12$ | $N=16$ | $N=8$ |
| 2.52 | 5.204 | 5.415 | 5.470 | 5.846 |
| 2.77 | 5.012 | 5.260 | 5.326 | 5.442 |
| 2.88 | 4.935 | 5.201 | 5.273 | 5.325 |
| 2.99 | 4.864 | 5.146 | 5.227 | 5.250 |
| 3.20 | 4.745 | 5.056 | 5.159 | 5.231 |

To get rid of lattice spacing errors we first work at fixed physical volume set by $M_WL = N\sqrt{\kappa/2}$ for three values of the lattice spacing $a = L/N$. We extrapolate $a \to 0$ by fitting the data to a linear function of $a^2$ (fig. 1). To check the stability of the fit we also performed a fit quadratic in $a^2$. In fig. 2 the linearly extrapolated values (errors determined by averaging intercepts computed from the three pairs of lattices) are plotted as a function of the physical size. Assuming that one can compute the fields in terms of the free massive propagator outside of the core region, one easily derives the volume dependence to be

$$\mathcal{E}_{\rm sph}(L) = \mathcal{E}_{\rm sph}^{\infty} + \gamma\, e^{-M_WL}(1 + \mathcal{O}(M_WL)^{-1})\ . \quad (9)$$

Using the extrapolations linear in $a^2$ we find for $\mathcal{E}_{\rm sph}^{\infty}$ a value of $5.02(1)M_W/\alpha_W$, dropping one (on which the fit in fig. 2 is based), two or three of the smallest volumes gives resp. 5.04(1), 5.05(1) and 5.06. On the other hand, using an extrapolation quadratic in $a^2$ gives resp. 5.01(3), 5.08(3), 5.12(3) and 5.16. As a conservative estimate we give $\mathcal{E}_{\rm sph}^{\infty} = 5.08(7)\, M_W/\alpha_W$, covering all values.

This work was supported in part by FOM and by a grant from NCF for use of the Cray C98. M.G.P. was also supported by MEC.

## REFERENCES

1. M. García Pérez and P. van Baal, Nucl. Phys. B429 (1994) 451.
2. L. Yaffe, Phys. Rev. D40 (1989) 3463; J. Kunz and Y. Brihaye, Phys. Lett. 216B (1989) 353.
3. R. Dashen e.a., Phys.Rev. D10 (1974) 4138.
4. F. R. Klinkhamer and N. Manton, Phys. Rev. D30 (1984) 2212.
5. G. Eilam and A. Stern, Nucl. Phys. B294 (1987) 775.
6. J. Ambjørn e.a., Nucl. Phys. B353 (1990) 346.
7. W. Langguth e.a., Nucl.Phys. B277 (1986) 11.